\documentstyle[aps,psfig,12pt,preprint]{revtex}
  
\begin{document}
\draft
\topmargin -0.5cm
\oddsidemargin 0in
\evensidemargin 0in
\textheight 8.5in
\textwidth 6.27in                                              

\title{Acoustic coupling between two air bubbles in water}
\author{\\ Pai-Yi HSIAO}
\address{
Laboratoire de Physique Th\'eorique de la Mati\`ere Condens\'ee \\
Universit\'e Paris 7 -- Denis Diderot \\
case 7020, 2 place jussieu, 75251 Paris Cedex 05, FRANCE\\
E-mail: $hsiao@ccr.jussieu.fr$
}
\author{\\ Martin DEVAUD and Jean-Claude BACRI}
\address{
Laboratoire des Milieux D\'esordonn\'es et H\'et\'erog\`enes \\
Universit\'e Pierre et Marie Curie -- Paris 6 \\
case 78, 4 place jussieu, 75252 Paris Cedex 05, FRANCE\\
E-mails: $devaud@ccr.jussieu.fr$, $jcbac@ccr.jussieu.fr$
}
\maketitle
\begin{abstract}
\\
{\bf Abstract --} The acoustic coupling between two air bubbles immersed
 in water is clearly demonstrated.
The system is acoustically forced, and its response is detected.
The experimental results confirm that both theoretically predicted eigenmodes,
respectively symmetrical and antisymmetrical, do exist.
Their frequencies, measured as a function of the bubbles spacing, follow
theoretical estimations within a  $10\%$  accuracy. 
\end{abstract}

\newpage

\section*{1. Introduction}
Bubbles play an important role in the sound propagation in everyday life
liquids. 
For example, the murmur of the brooks essentially originates, as first
suggested by Bragg\cite{minn,leighton}, in the oscillations of air bubbles captured and dragged
along by the water.
The so-called ``hot chocolate effect'', 
namely the rising of sound pitch when one
repeatedly taps the bottom of the mug in which some instant coffee or chocolate
is being dissolved, is explained by the releasing into the water of the tiny
bubbles trapped in the powder\cite{farrell,crawford}.
Bubble dynamics and acoustic properties of liquids containing a large number of
bubbles have been widely studied for a long
time\cite{devin,pros86,pros91,carst,silb,agos}.
A very good and complete review in this domain was achieved by
Leighton\cite{leighton2}.
{\it Inter alia}, the problem of the interaction of two neighbouring bubbles
has been discussed using fluid dynamics tools\cite{kuz} or the
acoustic-electrostatic analogy\cite{kob}. Moreover, the {\it free} 
oscillations of a
system of two (and even three) air cavities formed in a metal plate lying on a
water surface have been theoretically and experimentally investigated in detail
(including cubic nonlinearities)\cite{bre}. The aim of the present article is
to present a simple, readily reproducible, experimental study of the {\it
forced} oscillation regime of a two-air bubble system in water. We begin with a
short introductory theory in which we show that the two-bubble system is mostly
equivalent to a set of two magnetically coupled electric circuits.

\section*{2. Theoretical model}
An air bubble in water will be considered as a perfect sphere\footnote{
The correction to the Minnaert angular frequency due to deviation from the
spherical shape can be shown to be negligible\cite{stras0,robert,stras}.}
 of radius
$R(t)=R_0 + \xi(t)$, with variation $\xi$ much smaller than equilibrium value
$R_0$.
It can be shown that $\xi(t)$ oscillates with Minnaert's angular frequency
$\omega_0=\sqrt{3\gamma P_0/\rho_0R_0^2}$, where $\gamma$ is the specific heat
ratio $C_p/C_v$ of air, and $P_0$ and $\rho_0$ respectively stand for the
equilibrium pressure\footnote{
The pressure difference accross the bubble boundary due to air-water surface
tension is about $1\% P_0$ for a typical radius of $1mm$ (see \cite{devin}
eq.~(65b)) and will be neglected: $P_0$ is {\it also} the equilibrium pressure 
of enclosed air.}
 and mass density of water. 
This oscillation is damped through several mechanisms: of course the acoustic
radiation damping (thanks to which the bubble noise is audible), but also the
viscous and thermal dampings\cite{devin,pros91}. 
We will neglect, in the following simplified theory, the last two ones.
Moreover, allowing for the typical $1KHz$ acoustic frequency and
 $1mm$ bubble size we deal with 
in our experiment, we will neglect any sound propagation in the enclosed air.
We thus deliberately restrict the present study to the (radial) 
fundamental resonance of the air bubble-water system.

\subsection*{2.1 One-bubble free oscillation}
Let us consider one bubble with radius $R_0$ immersed in an infinite volume of
water at equilibrium pressure $P_0$. 
Let $P(\vec{r},t)$ be the actual pressure at site $\vec{r}$ and instant $t$.
The extra pressure $p(\vec{r},t)$ is defined as $P(\vec{r},t)-P_0$.
According to Minnaert's assumption, the enclosed air undergoes isentropic
transformations and its (extra) pressure $p(t)$ is homogeneous inside the
bubble.
Then, neglecting air's inertia as well as the air-water surface tension,
$p(t)$ and the radius variation $\xi(t)$ are linked by:
\begin{equation}
\frac{p(t)}{P_0}+\frac{3\gamma\xi(t)}{R_0}=0 \label{eq1}
\end{equation} 
On the other hand, it can be easily shown that (extra) pressure $p(r,t)$ at
distance $r$ from the center of the bubble follows a d'Alembert-like 1D
equation, 
the solution of which exactly reads, for $r\geq R_0$:
\begin{equation}
p(r,t) =\frac{1}{r}\rho_0R_0^2 \left[ \xi^{''}- \frac{R_0}{c} \xi^{'''}+ ...
       +(-\frac{R_0}{c})^k\xi^{(2+k)}+...\right](t-\frac{r-R_0}{c})
\end{equation}
where $c$ is the sound velocity in water.
If the acoustic wavelength $\lambda$ is much larger than $r$ (i.e., under the
circumstances, if condition $r\omega_0/c \ll 1$ is fulfilled), then
$p(r,t)$ can be approximated by:
\begin{eqnarray}
p(r,t) &\simeq& \rho_0 R_0^2 \left[\frac{\xi^{''}(t)}{r}-\frac{\xi^{'''}(t)}{c}
                               \right] \nonumber \\
       &\simeq&  \rho_0 R_0^2 \left[\frac{\xi^{''}(t)}{r}+\frac{\omega^2_0}{c}
                  \xi^{'}(t) \right] \label{eq3}
\end{eqnarray}
Then, equalling $p(t)$ in eq.~(\ref{eq1}) to $p(R_0,t)$ in eq.~(\ref{eq3}), one
gets, all calculations carried out:
\begin{equation}
\xi^{''}+\frac{\omega^2_0R_0}{c}\xi^{'}+\frac{3\gamma P_0}{\rho_0R_0^2}\xi =
\xi^{''}+\Gamma_{rad}\xi^{'}+\omega_0^2\xi=0 
\end{equation}
which is the well-known differential equation of a weakly\footnote{
Ratio $\Gamma_{rad}/\omega_0 = \omega_0 R_0 /c$ is actually assumed
to be much smaller than unity, as a consequence of the the validity condition
of eq.~(\ref{eq3}).} 
damped 1D harmonic oscillator.

\subsection*{2.2 Two-bubble free oscillation}
Let us now add a second bubble, with the same (equilibrium) radius $R_0$, at a
distance $d$ apart from the first one. Let $\vec{r_i}$ ($i=1,2$) be the
(equilibrium) position of the $i^{th}$ bubble center, $\xi_i(t)$ its radius
variation, $p_i(t)$ the (inner) extra pressure of the enclosed air, and 
$p_i(\vec{r},t)$ (resp. $\vec{u}_i(\vec{r},t)$ the would-be (outer)extra
pressure (resp. displacement with respect to equilibrium) at point $\vec{r}$
and instant $t$ in the water medium if bubble $i$ was alone.
Then, allowing for the superposition principle for small displacements, we
assume that overall water extra pressure and displacement respectively read:

\begin{eqnarray} 
      p(\vec{r},t) &=& p_1(\vec{r},t)+p_2(\vec{r},t) \label{eq5a} \\
\vec{u}(\vec{r},t) &=& \vec{u}_1(\vec{r},t) +\vec{u}_2(\vec{r},t) \label{eq5b} 
\end{eqnarray}

with, of course, $p_i(t)$ and $\xi_i(t)$ still linked by eq.(\ref{eq1}).
On the surface of the first bubble: 
$r_1 = \left|\vec{r}-\vec{r}_1\right| = R_0$,
$r_2 = \left|\vec{r}-\vec{r}_2\right| \simeq d$,  and
$p(\vec{r},t)=p_1(t)$.
A similar constraint is required on the surface of the second bubble, where
$r_1 \simeq d$ and $r_2 = R_0$. If the bubble spacing $d$ is much smaller
than $\lambda$ (i.e. $\omega_0 d/c \ll 1$)\footnote{
In our experiments $\lambda$ is of order $1m$, while $d$ ranges from $1$ to
$5\ cm$.},
 then eq.~(\ref{eq3}) is
available and we finally get the following pair of coupled motion equations:
\begin{eqnarray}
\xi_1^{''}+\alpha\xi_2^{''}+\Gamma_{rad}(\xi_1^{'}+\xi_2^{'})+\omega_0^2\xi_1=0
 \label{eq6a}\\
\alpha\xi_1^{''}+\xi_2^{''}+\Gamma_{rad}(\xi_1^{'}+\xi_2^{'})+\omega_0^2\xi_2=0 
 \label{eq6b}
\end{eqnarray}
where $\alpha=R_0/d\ (<0.5)$ can be regarded as a dimensionless coupling
constant.
Observe, by the way, that if double condition: $R_0\ll d \ll \lambda$ is
fulfilled, eqs.~(\ref{eq6a}) and (\ref{eq6b}) are available 
(with $\alpha \simeq 0$), and dynamic variables $\xi_i$ are still coupled by 
radiation
damping, since the dissipation terms do not involve $\alpha$.

Defining symmetrical and antisymmetrical normal variables $\phi_s(t)$  and
 $\phi_a(t)$ as respectively the sum and the 
difference of $\xi_1(t)$ and $\xi_2(t)$, we get the uncoupled equations system:
\begin{eqnarray}
(1+\alpha)\phi_s^{''}+2\Gamma_{rad}\phi_s^{'}+\omega_0^2\phi_s &=&0 
\label{eq7a}\\
(1-\alpha)\phi_a^{''}+\omega_0^2\phi_a &=&0 \label{eq7b}
\end{eqnarray} 
It is noteworthy that, as far as only radiation is concerned, the symmetrical
mode's damping rate is twice the single-bubble's one, while the antisymmetrical
mode is undamped.This feature is easily understood in terms of constructive
(resp. destructive) interference between the acoustic waves radiated by each
bubble, and parallels a well-known situation in the atomic physics domain
(super- and sub- radiant quantum states of a couple of identical atoms
interacting with each other through the E.M.~field).
From  eqs.~(\ref{eq7a}) and (\ref{eq7b}), it is clear that the symmetrical mode
has the lower angular frequency $\omega_s=\omega_0/\sqrt{1+\alpha}$, and the
antisymmetrical mode the higher one $\omega_a=\omega_0/\sqrt{1-\alpha}\ $.
Observe that, leaving apart the calculation of radiative damping, it is very
easy to derive above expressions of $\omega_{s,a}$ using the following trick.
Let us consider the water as an uncompressible fluid (i.e. $c \rightarrow
\infty$). The water displacement due to bubble $i$'s motion simply reads:
\begin{equation}
\vec{u}_i(\vec{r},t)=\xi_i(t)\frac{R_0^2}{r_i^2} \vec{e}_{r_i}  \label{radius}
\end{equation} 
with $\vec{e}_{r_i} = (\vec{r}-\vec{r}_i)/|\vec{r}-\vec{r}_i|
= (\vec{r}-\vec{r}_i)/r_i$. 
Then, allowing for eq.~(\ref{eq5b}), the
overall water kinetic energy $T$ is:
\begin{eqnarray}
T&=&\frac{1}{2}\rho_0\int d^3r\ \ (\frac{\partial\vec{u}}{\partial t})^2 
\nonumber\\ 
 &=&\frac{1}{2}M_0 \left( \xi_1^{'2}+ \xi_2^{'2}+2\alpha \xi_1^{'} \xi_1^{'}
\right) \label{eq9} 
\end{eqnarray}
where $M_0=4\pi R_0^3\rho_0$ is the effective mass of either bubble.
On the other hand, the total potential energy $V$ associated with the isentropic
compressibility of the enclosed air reads:
\begin{equation}
  V=\frac{1}{2} K \left(\xi_1^2 + \xi_2^2 \right)
\end{equation}
where $ K=12\pi \gamma R_0 P_0$ is the effective stiffness of either bubble.
The Lagrange equations derived from $L=T-V$ are:
\begin{eqnarray}
\xi_1^{''}+\alpha\xi_2^{''}+\omega_0^2\xi_1=0
 \label{eq11a}\\
\alpha\xi_1^{''}+\xi_2^{''}+\omega_0^2\xi_2=0
 \label{eq11b}                                                   
\end{eqnarray}
which is exactly the $c \rightarrow \infty$ limit of eqs.~(\ref{eq6a}) and
(\ref{eq6b}). 
It is worth noticing that eqs.~(\ref{eq9}) through (\ref{eq11b}) are formally
equivalent to those of a system of two ($L,C$) electric circuits coupled by
mutual induction with coefficient $\alpha L$. In this analogy, $M_0$ and $K$
respectively correspond to $L$ and $1/C$, and the $\xi_i$'s to the electric
charges $q_i$ of either capacitor. 

\subsection*{2.3 Forced oscillation}
Let us now suppose that the above studied two-bubble system is driven by an
external acoustic source with an angular frequency $\omega$ near Minnaert's one,
$\omega_0$. The phase difference of the driving pressures on both bubbles can
therefore be neglected, since $\omega d/c \ll 1$. Let $p_{ei}(t)$ be the
external pressure undergone by bubble $i$. Motion eqs.~(\ref{eq6a}) and 
(\ref{eq6b}) are then completed in:
\begin{eqnarray}
\xi_1^{''}+\alpha\xi_2^{''}+\Gamma_{rad}(\xi_1^{'}+\xi_2^{'})+\omega_0^2\xi_1
= -\frac{p_{e1}(t)}{\rho_0 R_0} \label{eq12a}\\
\alpha\xi_1^{''}+\xi_2^{''}+\Gamma_{rad}(\xi_1^{'}+\xi_2^{'})+\omega_0^2\xi_2
= -\frac{p_{e2}(t)}{\rho_0 R_0} \label{eq12b}
\end{eqnarray}                           
or equivalently:
\begin{eqnarray}
\phi_s^{''}+\frac{2\Gamma_{rad}}{1+\alpha}\phi_s^{'}+\omega_s^2\phi_s
&=& F_{es}(t) \label{eq13a}\\
\phi_a^{''}+\omega_a^2\phi_a &=& F_{ea}(t)\label{eq13b}
\end{eqnarray}                                
with $F_{es}(t)=-(p_{e1}(t)+p_{e2}(t))/\rho_0 R_0 (1+\alpha)$ and
     $F_{ea}(t)=-(p_{e1}(t)-p_{e2}(t))/\rho_0 R_0 (1-\alpha)$.
Solving for $\phi_s$ and $\phi_a$ in above eqs.~(\ref{eq13a}) and
(\ref{eq13b}), one gets $\xi_1(t)$ and $\xi_2(t)$, and consequently (using
eq.~(\ref{eq3})) quantities $p_1(\vec{r},t)$ and  $p_2(\vec{r},t)$ at any point
$\vec{r}$ of the medium. At last, comparing external (applied) pressure
$p_e(\vec{r},t)$ with the actual overall extra pressure 
$p(\vec{r},t)=p_e(\vec{r},t)+p_1(\vec{r},t)+p_2(\vec{r},t)$, we can
experimentally measure the two-bubble system's response as a function of
$\omega$.
In this respect (and provided that the excitation-detection geometry allows
it), resonances are expected for $\omega=\omega_s$ and $\omega=\omega_a$.

\section*{3. Experiments}
Our aim is to demonstrate the existence of both above mentioned modes.
From an experimental point of view, it turns out to be easier to implement a
forced oscillation scheme than a free oscillation one. We therefore present the
former hereafter.
\subsection*{3.1 Experimental setup}
A small net (see fig.1), made up with a gauze maintained with a wire, is
designed to catch up an air bubble in water and to fix it at any desired
position without appreciably modifying acoustic impedance
and spherical symmetry.
Two such devices are used for studying the two-bubble system. The external
driving source is a speaker and extrapressure $p(\vec{r},t)$ is measured with a
small microphone. A function generator, to which the speaker is connected,
produces a c.w.~sinusoidal signal with a frequency slowly swept from
$f_{low}$ to $f_{high}$. The signal delivered by the microphone is transmitted
to a lock-in amplifier which compares it with the reference one (delivered by
the function generator) and decomposes it into real and imaginary parts. Both
parts can be seen on an oscilloscope and recorded with a computer (see fig.2).

In a preliminary set of experiments, without any bubble in the aquarium, the
response of the microphone is calibrated for different speaker-microphone
configurations.
Two kinds of configurations are presented in figure~3.
In figs.~3(a) and 3(b) the configuration is deliberately asymmetrical:
 the microphone
is mainly susceptible to bubble 2's motion, while the speaker selectively
drives bubble 2 (fig.~3(a)) or bubble 1 (fig.~3(b)),
 so that $F_{ea}(t)$ is nonzero:
both modes can be excited and the associated motions detected.
 In fig.~3(c), the speaker is placed far from the bubbles; then, not only the
phases, but also the amplitudes of the external pressures $p_{1e}(t)$ and 
$p_{2e}(t)$ undergone on either bubble are appreciably equal. In such a
symmetrical excitation configuration, $F_{ea}(t) = 0$, so that the
antisymmetrical mode remains unexcited. Observe, by the way, that since
distances $r_1$ and $r_2$ between the bubbles and the microphone are equal, the
latter would detect no contribution from the antisymmetrical mode
 {\it even though} it was excited
(see eqs.(\ref{eq3}) and (\ref{eq5a}): $r_1 = r_2$ and $\xi_1=-\xi_2$ yields
$p_1(r_1,t)+p(r_2,t) = 0$ ). 

\subsection*{3.2 Results and discussion}
In figs.~4(a) and 4(b), the imaginary part $Im\,p$ of the output signal from the
lock-in amplifier is displayed versus the speaker frequency $f$ for 
various values of the bubbles spacing $d$.  
Figures 4(a) and 4(b) respectively correspond to configurations 3(a) and 3(b).
Two resonances can be made out in fig 4(a) and (though at a lesser degree) in
fig 4(b). Observe that the sign of the signal at the higher frequency resonance
is changed from 4(a) to 4(b), while the lower frequency one remains unchanged.
This is consistent with the latter signal being associated with the symmetrical
mode's resonance ($\omega_s = \omega_0/\sqrt{1+\alpha} < \omega_0$, and
$F_{es}$ unchanged from configuration 3(a) to 3(b)), and the former one with
the antisymmetrical mode's resonance $(\omega_a= \omega_0/\sqrt{1-\alpha}>
\omega_0$, and $F_{ea}$ changed into $-F_{ea}$ from configuration 3(a) to 3(b)).

It is noteworthy that both resonances have appreciably the same width. This is
in contradiction with simplified eqs.(\ref{eq7a}) and (\ref{eq7b}) (or
(\ref{eq13a}) and (\ref{eq13b})), in which only the radiation damping was
considered. In fact, as mentioned in introduction, other kinds of damping
should be taken into account: if viscous damping is absolutely negligible
for such
large bubbles, thermal damping is not (see fig.8 in \cite{pros77}),
and may be at the origin of the linewidth.
 Further discussion of 
this point is out of the scope of the present paper.
In figure 5, we have plotted, for both symmetrical and antisymmetrical modes,
and for $R_0 \simeq 2mm$, the inverse squared frequency $f^{-2}$ (multiplied by
a factor of $10^7$) versus the inverse bubble spacing $d^{-1}$, in order to get
a visual check of theoretical relations:
\begin{eqnarray}
\frac{1}{f^2_s}=\frac{1}{f^2_0}+\frac{R_0}{f_0^2}\cdot\frac{1}{d} \\
\frac{1}{f^2_a}=\frac{1}{f^2_0}-\frac{R_0}{f_0^2}\cdot\frac{1}{d} 
\end{eqnarray}

Although experimental points are appreciably
aligned, the measured slopes are about $40\%$ below theoretical prediction,
suggesting that coupling constant $\alpha$ has been overestimated. In fact,
theoretical value $\alpha = R_0/d$ was derived in eq.(\ref{eq9}) when
integrating the water kinetic energy density $\frac{1}{2}\rho_0(\frac{\partial 
\vec{u}}{\partial t})^2$
 over the whole space\footnote{
More precisely: over the whole space {\it outside} the two bubbles
 (the inner air's inertia being negligible). 
Nevertheless, it can be shown that the coefficient
of coupling term $\xi_1^{'} \xi_2^{'}$ in integral (\ref{eq9})
{\it does not} depend
 on the bubbles radius $R_0$.}. 
This inertial coupling is naturally lowered if some obstacle lies between the
bubbles and consequently screens (part of) the water flow\footnote{
The effective mass $M_0$ is modified too, but at a lesser degree.}.
Now, this is exactly what happens in configurations 3(a) and 3(b): to be able
to excite the antisymmetrical mode, we are compelled to insert the speaker
between the two bubbles, thus bringing about the above screening effect. In
order to check this interpretation, we performed the same experiment with
configuration 3(c), and recorded, for the symmetrical branch of the linear
fitting of fig.~5, a
slope of about $90\%$ of the theoretically predicted value.

At last, it should be noted that, when blocked from the top by the net,
the bubble is, strictly speaking, no longer spherical. 
As mentioned in footnote 1, such a deviation from the spherical shape is
(almost) of no consequence, and we have used the formulas derived above in
section 2 with $R_0$ standing for the radius of the sphere of equivalent
volume ({\it i.e.} the radius of the bubble before it is captured by the net).
Nevertheless, this feature of our experimental setup raises the following
difficulty: since, in course of motion, the fixed point of the bubble is no
longer its centre (as implicitly assumed in the theoretical model) but
its top, expression (\ref{radius}) of the water displacement is no longer
correct; a dipolar term should be added to the spherical monopolar one.
As a consequence, the kinetic energy $T$ derived in eq.(\ref{eq9})
 is modified too.
 More precisely, an exact calulation shows that $T$ should be multiplied
by a factor of $7/6$ and the coupling constant $\alpha$ by a factor of
$1 + R_0^2/(4d^2)$. 
These corrections lie within our experimental accuracy. We have consequently
neglected them. In this respect, it may be noted that the gauze in our device 
{\it does not} act like a rigid wall because the water can flow through the
meshes of the net. The situation is therefore different from that discused
in other studies\cite{stras0,howkins} considering the influence of the
proximity of a rigid boundary on the Minnaert frequency.

\section*{4. Conclusion}
As a conclusion, the acoustic inertial coupling between two air bubbles in
water has been experimentally put in evidence. Theoretical analysis shows that
the two-bubble system is formally equivalent to a set of two magnetically
coupled ($L$,$C$) electric circuits, with two eigenmodes, respectively
symmetrical and antisymmetrical. Experimental measurements and theoretical
predictions are in $10\%$ accuracy agreement.

\section*{5. Acknowledgment}
We gratefully thank Professor F.Massias for enriching discussions about
the correction to the Minnaert frequency in the case of a bubble fixed from
the top by a wide-mesh net.


\newpage
\section*{Figure Captions}
\begin{description}

\item[FIG.1]
 Simple tool for capturing the bubble

\item[FIG.2]
 Diagram of the experimental setup.
 In the experiment, we pump air into a tube immersed in water to produce
 the bubbles. The radii difference between these bubbles is small and
 will be neglected. (It can be shown that a small radii difference
 yields second order correction of $\omega_s$ and $\omega_a$).           

\item[FIG.3]
 Different geometrical configurations

\item[FIG.4]
 {\bf (a)} Spectra of symmetrical and antisymmetrical modes in
 configuration~3(a). \\
 {\bf (b)} Change of sign of $Im\, p$ for the antisymmetrical mode when
 configuration~3(b) is adopted.                                         

\item[FIG.5]
 Linear fitting of the plot $\frac{1}{f^2_{s,a}}\cdot 10^7 (Hz^{-2})$
 vs. $\frac{1}{d} (cm^{-1})$ for the two modes. Average of the resonance
 frequency of the two bubbles: $1499 Hz$;  corresponding radius: $0.217 cm$;
 slopes for the two fitting lines: $0.578$ and $-0.550$ $(cm\cdot sec^2)$;
 slopes of theoretical prediction: $\pm 0.966 (cm\cdot sec^2)$.        

\end{description}

\newpage

\begin{figure}
\centerline{\psfig{file=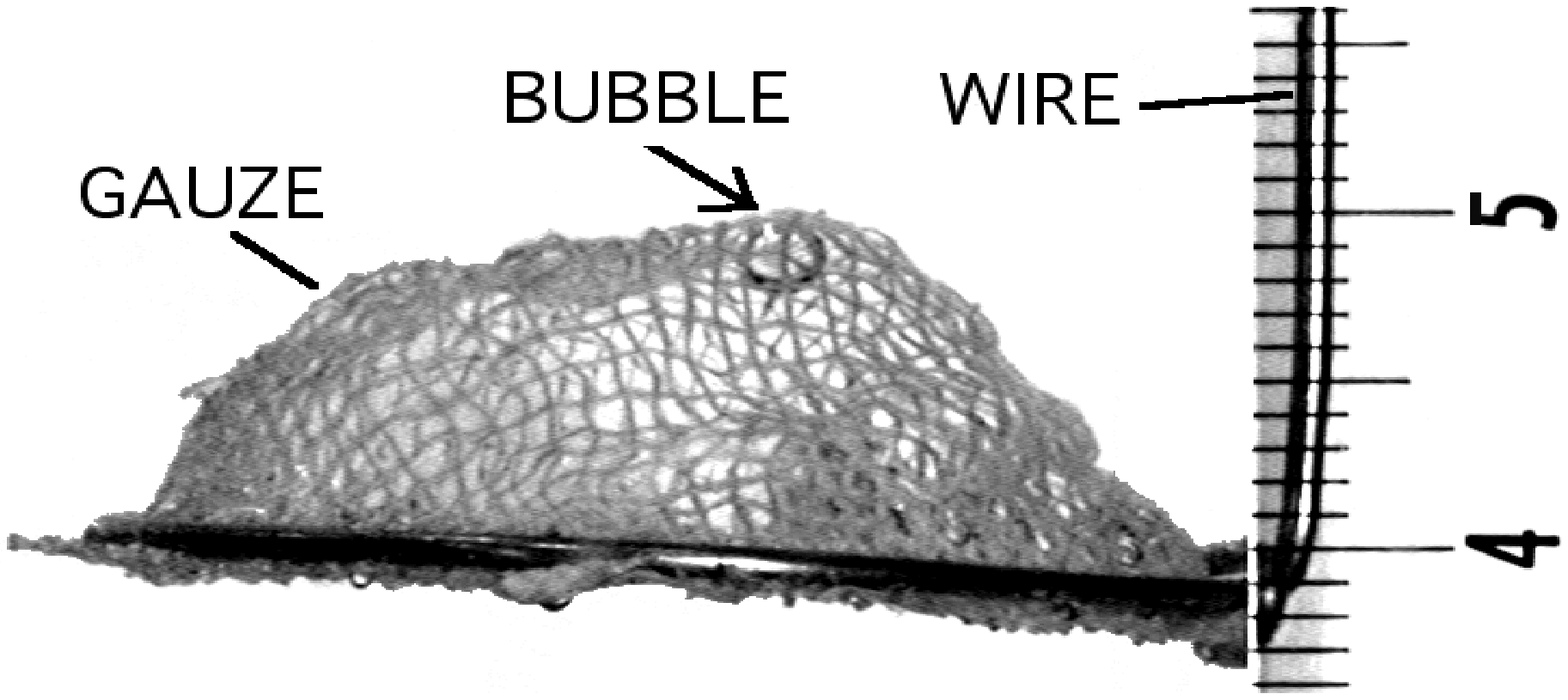}}
\center{\bf FIG. 1}
\end{figure}                         

\begin{figure}
\centerline{\psfig{file=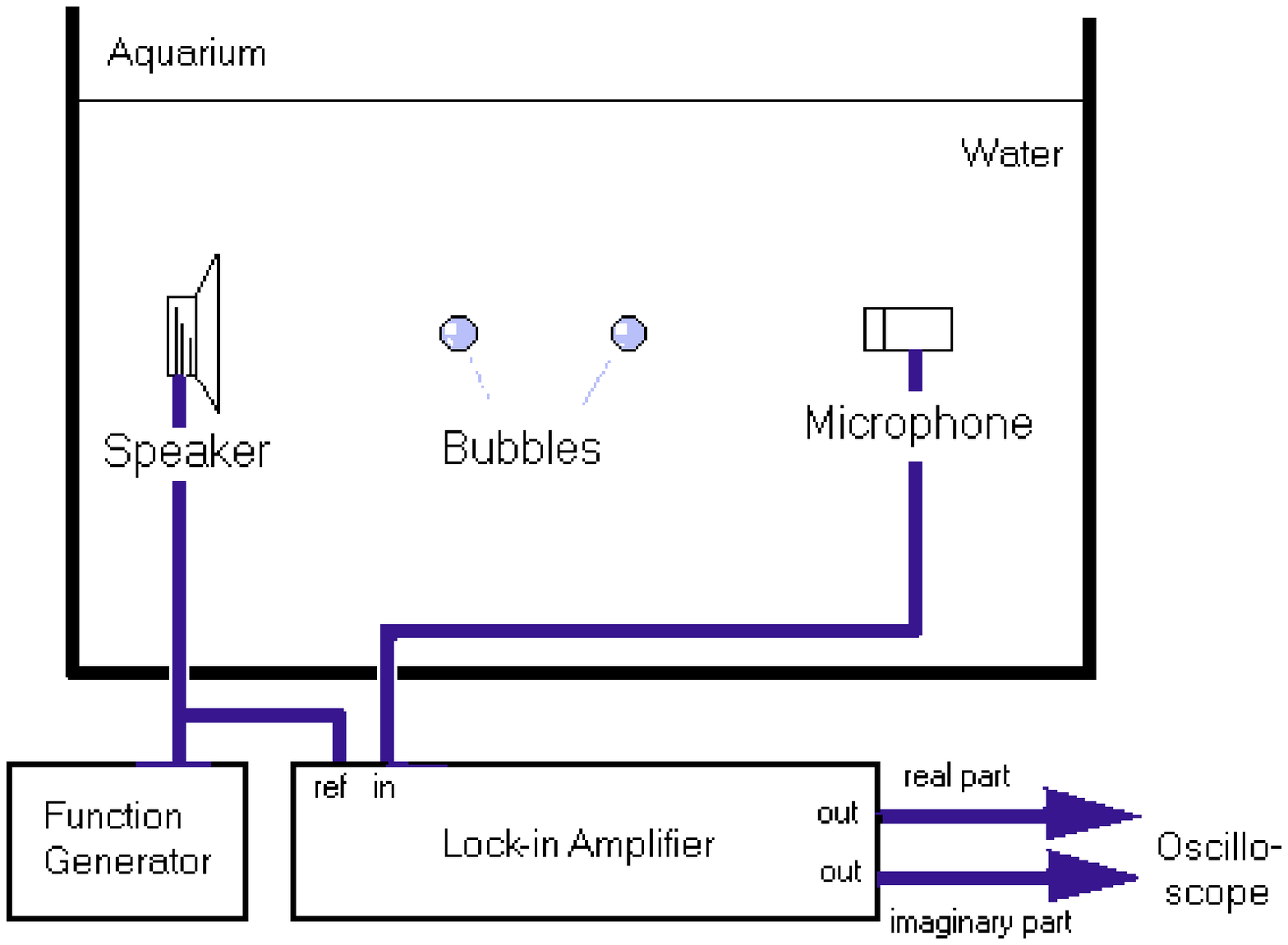}}
\center{\bf FIG. 2} 
\end{figure}                    

\begin{figure}
\centerline{\psfig{file=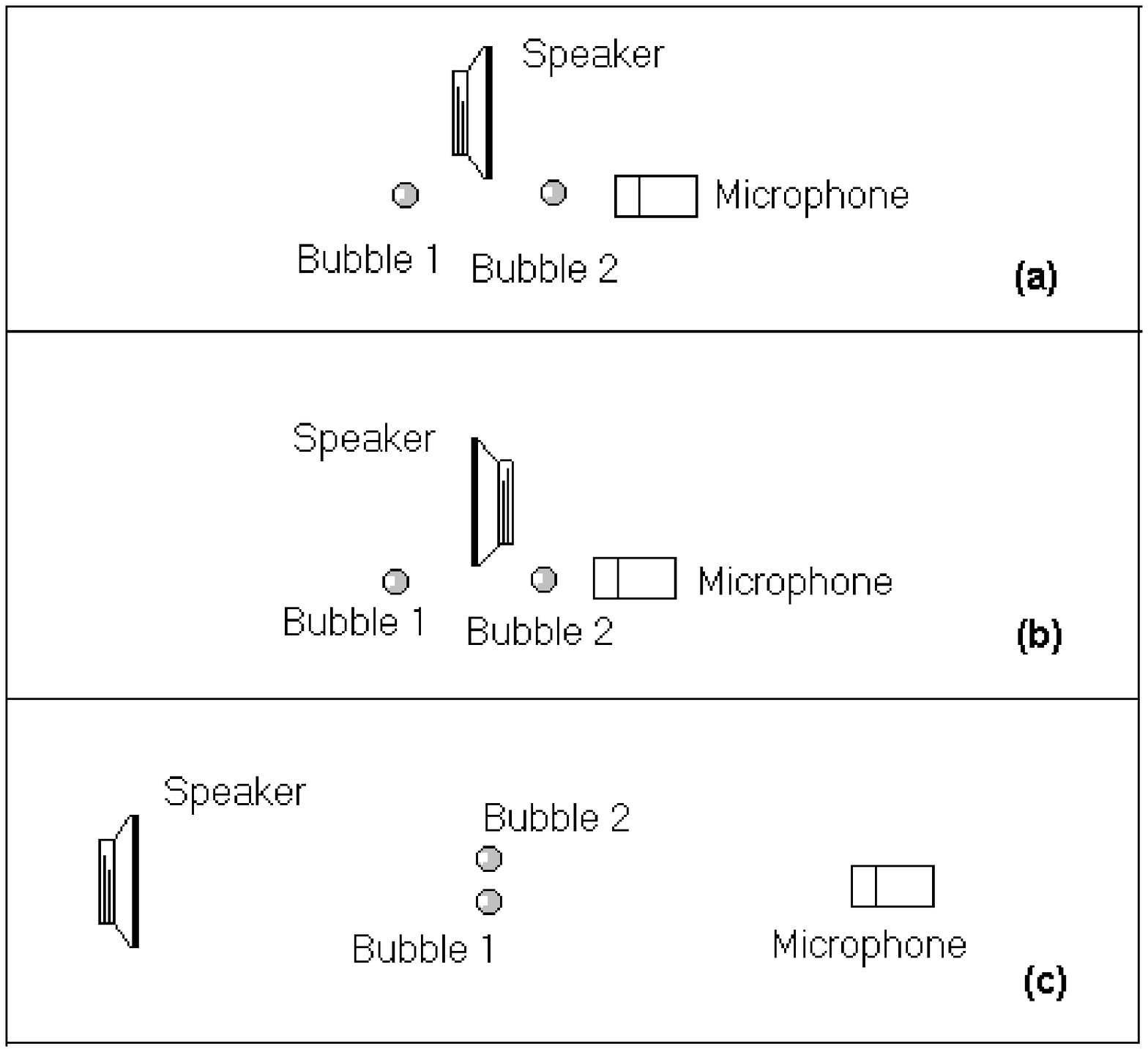}}
\center{\bf FIG. 3} 
\end{figure}                

\begin{figure}
\centerline{\psfig{file=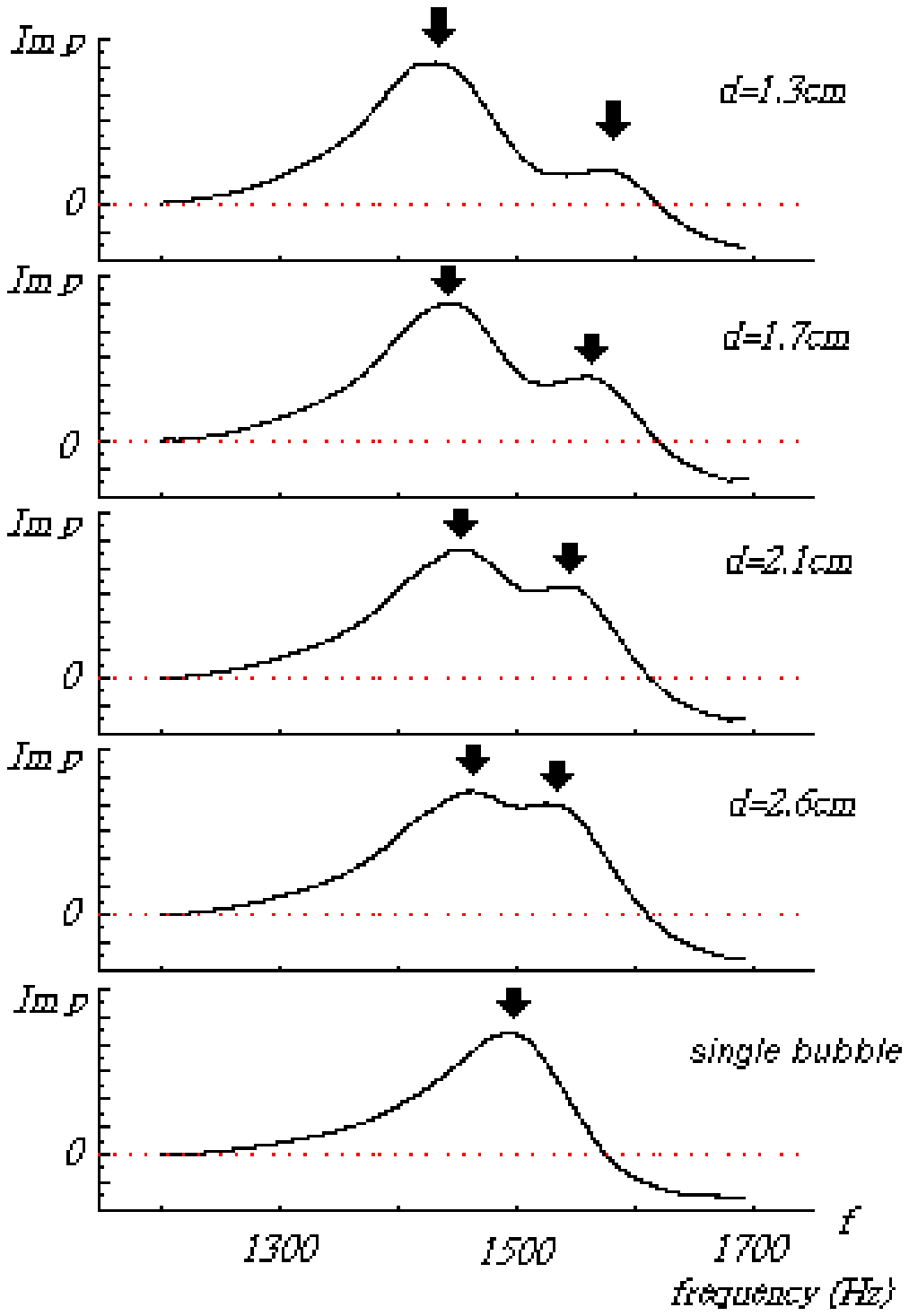}}
\center{\bf FIG. 4(a)} 
\end{figure}                  

\begin{figure}
\centerline{\psfig{file=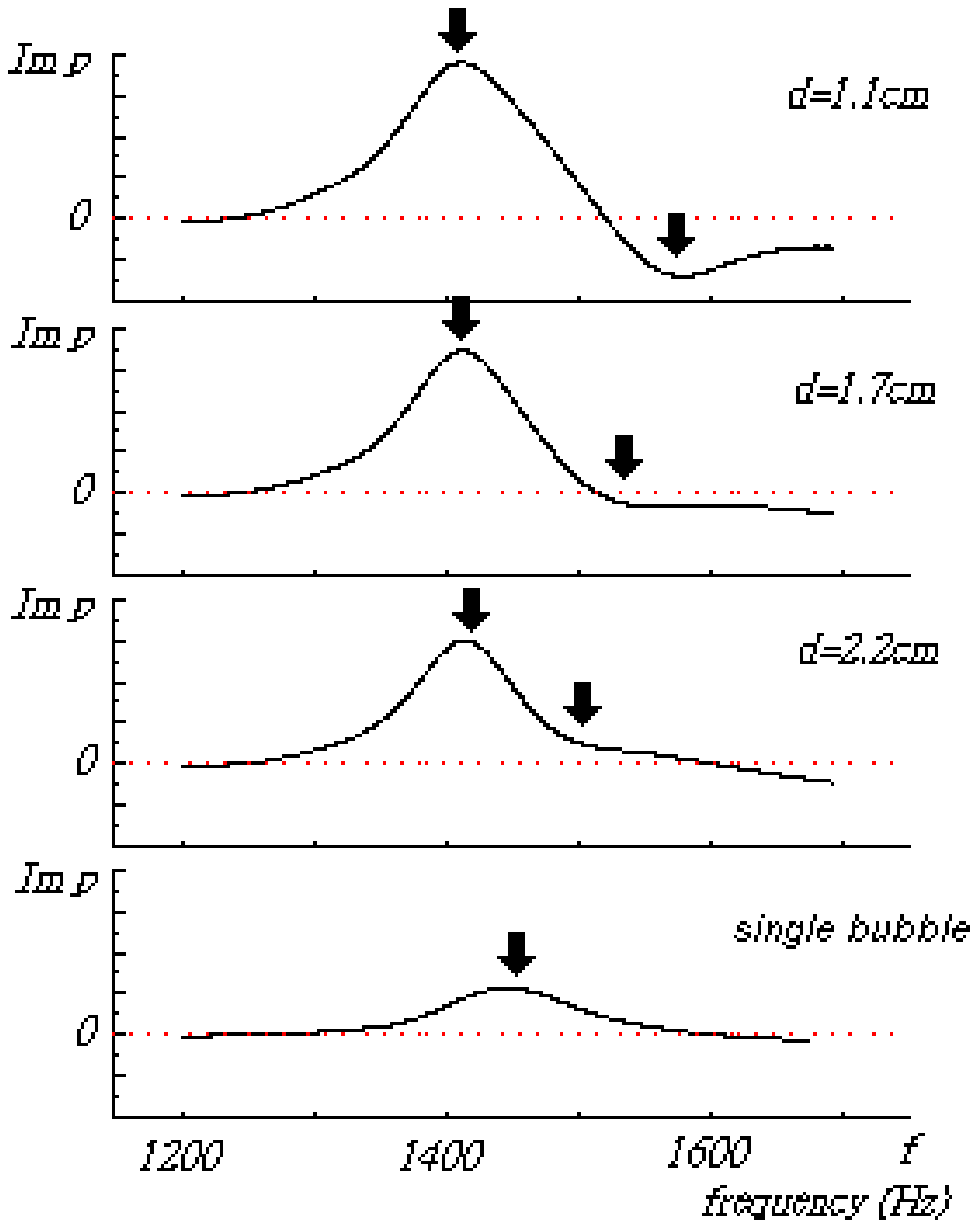}}
\center{\bf FIG. 4(b)}
\end{figure}                      

\begin{figure}
\centerline{\psfig{file=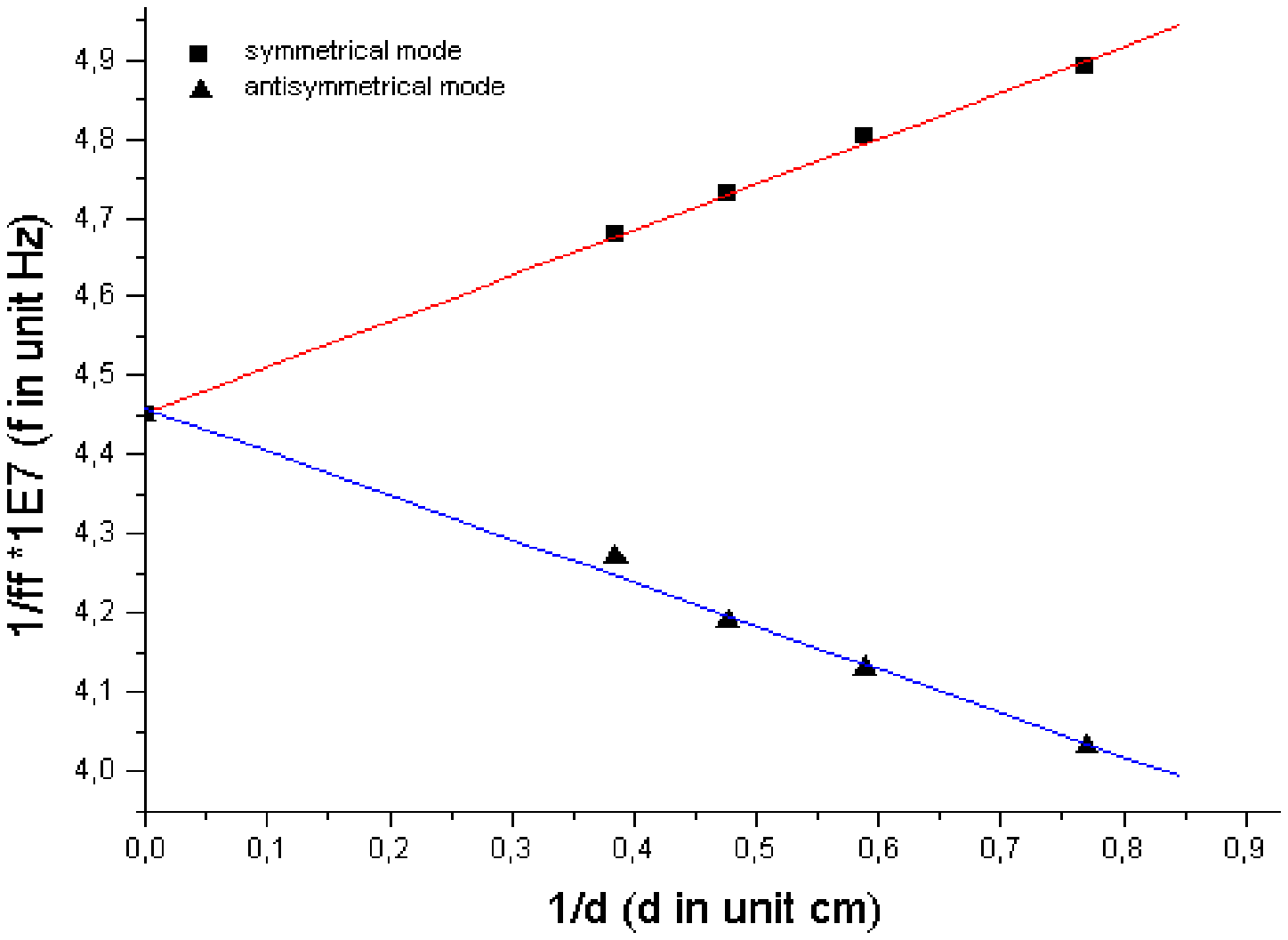}}
\center{\bf FIG. 5} 
\end{figure}    

\end{document}